\documentclass[runningheads]{llncs}
\usepackage{graphicx}
\usepackage{todonotes}
\usepackage[acronym,shortcuts]{glossaries}
\usepackage{amsmath}
\usepackage{amsfonts}
\usepackage[capitalize]{cleveref}
\def\R{\ensuremath{\mathbb{R}}}

\glsdisablehyper
\newacronym{wsi}{WSI}{whole slide image}
\newacronym{msi}{MSI}{microsatellite instability}
\newacronym{ssl}{SSL}{self-supervised learning}

\begin{document}
\title{Joint multi-task learning improves weakly-supervised biomarker prediction in computational pathology}
\titlerunning{Joint multi-task learning in computational pathology}
\author{Omar S.~M.~El Nahhas\inst{1}\and
Georg W\"{o}lflein \inst{2} \and
Marta Ligero \inst{1} \and
Tim Lenz \inst{1} \and
Marko van Treeck \inst{1} \and
Firas Khader \inst{3} \and
Daniel Truhn \inst{3} \and
Jakob Nikolas Kather \inst{1,4, 5}}

% \author{anonymous\inst{1}\and
% anonymous \inst{2} \and
% anonymous \inst{1} \and
% anonymous \inst{1} \and
% anonymous \inst{1} \and
% anonymous \inst{3} \and
% anonymous \inst{3} \and
% anonymous \inst{1,4, 5}}
%
\authorrunning{El Nahhas et al.}
% \authorrunning{anonymous}
\institute{Else Kroener Fresenius Center for Digital Health, Medical Faculty Carl Gustav Carus, TUD Dresden University of Technology, Germany \and
School of Computer Science, University of St Andrews, St Andrews, United Kingdom \and
Department of Diagnostic and Interventional Radiology, University Hospital Aachen, Aachen Germany \and
Department of Medicine 1, University Hospital and Faculty of Medicine Carl Gustav Carus, TUD Dresden University of Technology, Germany \and
Medical Oncology, National Center for Tumor Diseases (NCT), University Hospital Heidelberg, Heidelberg, Germany
}
% \institute{anonymous organization \\ anonymous organization \and
% anonymous organization \\ anonymous organization \and
% anonymous organization \\ anonymous organization \and
% anonymous organization \\ anonymous organization \and
% anonymous organization \\ anonymous organization
% }
%
\maketitle             % typeset the header of the contribution
%
% \todo[]{anonymize (also don't forget to remove acknowledgements)}
\begin{abstract}
Deep Learning (DL) can predict biomarkers directly from digitized cancer histology in a weakly-supervised setting. Recently, the prediction of continuous biomarkers through regression-based DL has seen an increasing interest. Nonetheless, clinical decision making often requires a categorical outcome. Consequently, we developed a weakly-supervised joint multi-task Transformer architecture which has been trained and evaluated on four public patient cohorts for the prediction of two key predictive biomarkers, microsatellite instability (MSI) and homologous recombination deficiency (HRD), trained with auxiliary regression tasks related to the tumor microenvironment. Moreover, we perform a comprehensive benchmark of 16 task balancing approaches for weakly-supervised joint multi-task learning in computational pathology. Using our novel approach, we outperform the state of the art by +7.7\% and +4.1\% as measured by the area under the receiver operating characteristic, and enhance clustering of latent embeddings by +8\% and +5\%, for the prediction of MSI and HRD in external cohorts, respectively.

\end{abstract}

\begin{figure}[!ht]
    \centering
    \includegraphics[width=\textwidth]{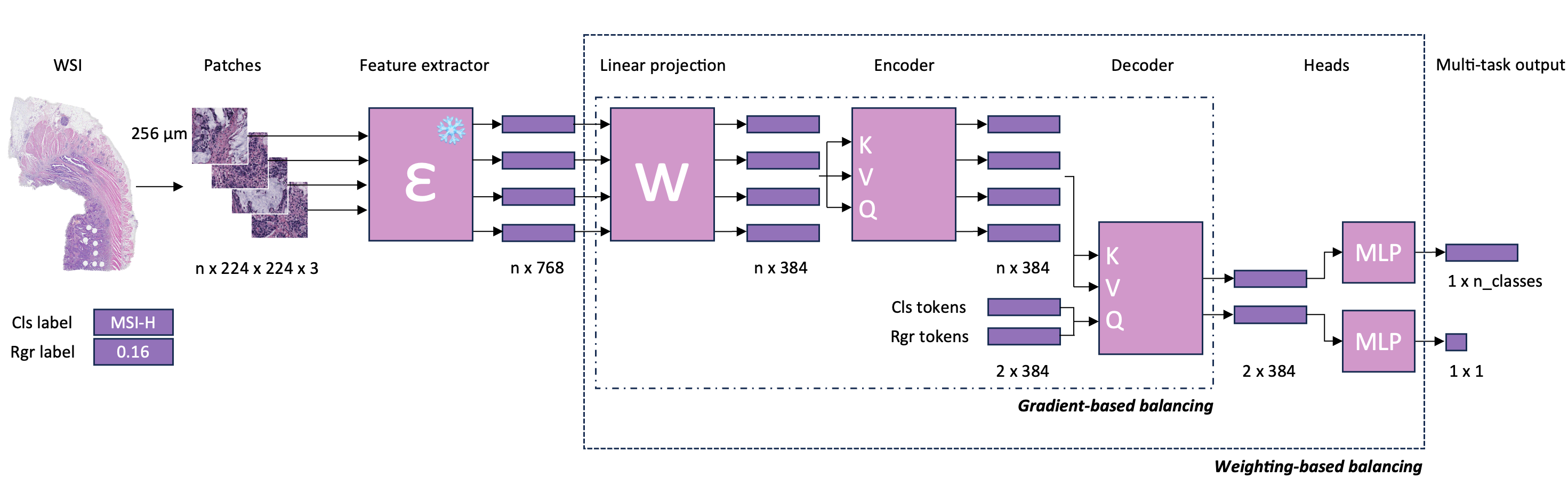}
    \caption{Model overview. We tessellate \glspl{wsi} into patches, extract CTransPath features~\cite{Wang2022-wg}, linearly project them, and feed them into a Transformer encoder. A learnable classification and regression token are added to the input of the Transformer decoder, after which the output is fed to a classification and regression head, performing weakly-supervised joint multi-task learning with weighting- and gradient-based task balancing.}
    \label{fig:overview}
\end{figure}

\section{Introduction}
Over the past years, Deep Learning (DL) has proven its utility in predicting biomarkers directly from \glspl{wsi} with hematoxylin- and eosin (H\&E)-staining in a weakly-supervised manner. Weakly-supervised learning in computational pathology allows for large-scale analyses using solely the reported diagnosis as training labels, eliminating the need for cost- and time expensive pixel-level annotations~\cite{Campanella2019-ni}. The majority of studies predict categorical biomarkers with classification-based methods~\cite{Kather2019-jh,Wagner2023-zn}, with a recent study showing the benefit of applying a regression-based method instead of dichotomizing the target for reformulation as a classification problem~\cite{El_Nahhas2024-xu}. The studies predominantly follow the same pattern for model validation, often using heatmaps, top tiles and concordance analyses to confirm the model's alignment with known biological concepts~\cite{Niehues2023-ew,Loeffler2023-vt}. For example, biomarkers such as microsatellite instability (MSI) and homologous recombination deficiency (HRD) are predictive biomarkers which have known correlations with immune cells in the tumor microenvironment (TME)~\cite{Bai2021-wg,Shi2023-pf}. However, the current state of the art for predicting MSI and HRD do not use observations from the tumor microenvironment as an additional learned task~\cite{Wagner2023-zn,El_Nahhas2024-xu}, potentially leaving room for improved biomarker prediction. This leads to our primary research question: \textit{Does including additional biological information in the form of an auxiliary regression task improve the prediction performance of the main classification task in weakly-supervised computational pathology?} Consequently, we develop and evaluate a joint multi-task learning Transformer model which focuses on predicting the main classification task of MSI or HRD, while learning additional information about the TME through an auxiliary regression task in a weakly-supervised setting. 

Our contributions are as follows:
\begin{enumerate}
    \item We propose a weakly-supervised joint multi-task learning framework that allows for additional biological information about the tumor microenvironment to be learned to improve the main biomarker prediction objective.
    \item We conduct the first comprehensive benchmark of 16 multi-task balancing approaches in weakly-supervised computational pathology.
    \item We improve over state-of-the-art weakly-supervised classification models for 2 highly relevant biomarkers, MSI and HRD, in 4 publicly available cohorts. Furthermore, we publicly release our code to promote reproducibility.
\end{enumerate}

\section{Related work}
The concept of multi-task learning has been applied to the field of computational pathology for H\&E \glspl{wsi} in various studies. Yan et al.~\cite{Yan2020-dg} and Graham et al.~\cite{Graham2023-iv} combined segmentation and classification tasks using cross-entropy (CE) losses which are summed with equal weights for each task. A variety of studies combined solely classification objectives in a multi-task setting, using CE losses which are equally summed across the tasks~\cite{Tellez2020-gz,Mormont2020-zs}, or as a weighted sum with constants found through a hyperparameter search~\cite{Lu2021-pp,Marini2021-em}. Gao et al.~\cite{Gao2023-no} combined a CE loss with a mean-squared error (MSE) loss which are balanced according to preset constants which only update in specific, pre-defined scenarios, and are manually bounded. Only Lu et al.~\cite{Lu2021-pp} and Marini et al.~\cite{Marini2021-em} approached the multi-task problem from a weakly-supervised perspective.
% conclusion
In summary, prior studies opted for weighted-based balancing approaches for multi-task learning, which were either equally balanced, or fine-tuned for very specific use-cases that likely do not translate well to other scenarios of a similar kind~\cite{Tellez2020-gz}. This leaves a clear gap in the computational pathology literature for the application of more sophisticated, model-guided balancing of losses and gradients~\cite{Chen2020-fd,Kendall2017-vx,Liu2021-cs,Liu2022-ao,Liu2018-rh,Loshchilov2018-uv,Yu2020-wy}, especially in a weakly-supervised setting.

\section{Method}
We consider a dataset of $N$ \glspl{wsi} $\mathbf{X}^{(1)},\dots,\mathbf{X}^{(N)}$, where each \gls{wsi} $\mathbf{X}^{(i)}\in\R^{W \times H \times 3}$ is an RGB image of width $W$ and height $H$, though these dimensions may vary between slides. 
During training, each \gls{wsi} $\mathbf{X}^{(i)}$ is associated with a binary
%\footnote{We consider binary classification in this paper, but note that our method can be straightforwardly extended to multiclass classification problems.} 
classification label $y^{(i)}\in\mathcal{Y}=\{0,1\}$ for the main task, as well as an auxiliary regression label $a^{(i)}\in\R$.
For example, the classification label $y^{(i)}$ could indicate MSI status, and the auxiliary target $a^{(i)}$ could represent a molecular signature for lymphocyte infiltration, which takes on continuous values.

Due to their large size, it is common to consider \glspl{wsi} as collections of patches, framing the \gls{wsi} classification problem as a weakly supervised learning task. 
More specifically, we split each \gls{wsi} $\mathbf{X}^{(i)}$ into a set of $n$ non-overlapping patches $\{\mathbf{x}^{(i)}_1,\mathbf{x}^{(i)}_2,\dots,\mathbf{x}^{(i)}_n\}$ where each $\mathbf{x}^{(i)}_j\in\mathcal{X}=\R^{P\times P\times 3}$ for a fixed patch size $P$ (the number of patches $n$ varies depending on the particular slide's dimensions).
We follow the STAMP protocol~\cite{El_Nahhas2023-du}, which sets the patch size $P=224$ at an edge length of 256 microns (which corresponds to approximately $9\times$ magnification), yielding $n^{(i)}\in \mathbb{N}$ non-background patches per slide.
The task is to train a model $M : \mathcal{P}(\mathcal{X})\to\mathcal{Y}$ that at inference time predicts the classification label given a bag of patches representing a \gls{wsi}.
During training, this model should learn from both the classification labels $y$ and the auxiliary regression target $a$, though at inference time we are only interested in the former. 

Obtaining the prediction from a collection of patches representing a \gls{wsi} is a two-step process consisting of (i) feature extraction and (ii) feature aggregation, outlined in \cref{fig:overview}. 
We describe these steps in the sections below. The source code is available at: \url{https://github.com/Avic3nna/joint-mtl-cpath}.

\subsection{Feature extraction}
\label{sec:feature_extraction}
Our model operates on feature vectors instead of raw patches.
Thus, we first apply a feature extractor $\mathcal{E}: \mathcal{X} \to \mathbb{R}^{d_z}$ individually to each patch $\mathbf{x}^{(i)}_j$ in order to obtain a corresponding feature vector $\mathbf{z}^{(i)}_j=\mathcal{E}\left(\mathbf{x}^{(i)}_j\right)$ that meaningfully represents each patch.
We parameterize $\mathcal{E}$ with CTransPath~\cite{Wang2022-wg}, a model that was pretrained on 32,000 \glspl{wsi} across various cancer types using self-supervised learning.
The extracted CTransPath feature vectors, which are of dimensionality $d_z=768$, are cached before training to save compute. 
As such, our preprocessing and feature extraction setup closely follows the STAMP~\cite{El_Nahhas2023-du} protocol.
However, unlike STAMP, we do not perform stain normalisation because a recent study found no effects of stain normalization on CTransPath feature embeddings, whilst incurring substantial computational overhead~\cite{Wolflein2023-vp}.

\subsection{Architecture}
\label{sec:architecture}
The proposed joint multi-task Transformer architecture (\cref{fig:overview}) is a modified version of the one found in Vaswani et al.~\cite{Vaswani2017-ht}: First, we project the features into a lower-dimensional latent space, to prevent the Transformer architecture's complexity from exploding for high-dimensional features. We then encode these projected input tokens using a Transformer encoder stack. Next, we decode these tokens using \texttt{[cls]} tokens for the main classification task alongside additional \texttt{[rgr]} tokens for the auxiliary regression task. Finally, we forward each of the decoded tokens through a fully connected layer to get a label-wise prediction. We opted for this architecture instead of the classic Vision Transformer \cite{Dosovitskiy2020-gr} to improve performance for multi-task, multi-label predictions that can scale across many tasks. Specifically, the architecture differs from the one proposed by Vaswani et al.~\cite{Vaswani2017-ht} in 1) an initial projection stage that reduces the dimension of the feature vectors and enables using the Transformer with larger input feature dimensions and 2) a set of fixed, learned class tokens in conjunction with equally as many independent fully connected layers to predict multiple labels at once. %instead of their language translation task, where the tokens of the words translated so far are used to predict the next word in the sequence. Our empirical experiments showed that adding too many class tokens to a Vision Transformer decreases its performance, as the same weights have to both process the tiles' information as well as the class tokens. Using our architecture alleviates these issues, as the data-flow of the class tokens is completely independent of the encoding of the tiles.

\subsection{Training}
All models are trained in a weakly-supervised setting. All experiments performed within the scope of this study use CTransPath feature vectors~\cite{Wang2022-wg}, are trained using 5-fold cross-validation with an 80-20 split for training and testing, and have the exact same patient split for each fold across all the compared models. The area under the receiver operating characteristic (AUROC, AUC), area under the precision-recall curve (AUPRC, PRC), and silhouette score (SS) are reported with the mean of the 5-folds of each experiment. The baseline model performs solely classification on the main task of predicting MSI or HRD, whereas the joint-learned model additionally performs regression on the auxiliary task of predicting a tumor microenvironment signature such as lymphocyte infiltrating signature score (LISS), leukocyte fraction (LF), stromal fraction (SF), tumor cell proliferation (Prolif), and intratumor heterogeneity (ITH). The model is optimized using AdamW \cite{Loshchilov2018-uv} with a learning rate of 1e-4, with the CE loss and MSE loss for classification and regression, respectively. The batch size is 1, using all \textit{n} patch features for each \gls{wsi} during training, for 32 epochs. Early stopping is triggered when the CE loss of the primary classification task shows no decrease for 7 consecutive epochs.

\subsection{Multi-task balancing}
We apply and compare a total of 16 task balancing approaches for the joint multi-task learning experiments. For weighting-based balancing, we use uncertainty (\textit{uncert})~\cite{Kendall2017-vx}, dynamic weight averaging (\textit{dwa})~\cite{Liu2018-rh}, and Auto-Lambda (\textit{autol})~\cite{Liu2022-ao}. For gradient-based balancing, we use gradient sign dropout (\textit{graddrop})~\cite{Chen2020-fd}, projecting conflicting gradients (\textit{pcgrad})~\cite{Yu2020-wy}, and conflict-averse gradient descent (\textit{cagrad})~\cite{Liu2021-cs}. For comparison with methods used in prior studies, we include an approach which weights the tasks equally (\textit{naive}). Previous work states that combining weighting- and gradient-based balancing in multi-task learning can improve performance~\cite{Liu2022-ao}, which leads to the combination of aforementioned methods. All balancing approaches focus on single objective optimization, i.e. improving the classification performance regardless of the regression performance, except for \textit{autol} which performs multi-objective optimization for both classification and regression~\cite{Liu2022-ao}. Weighting-based balancing affects all non-frozen layers in the network, whereas gradient-based balancing only affects the shared projector, encoder and decoder layers (Fig.~\ref{fig:overview}).

\section{Experiments and Results}
\subsection{Data}
We use four public cohorts for the training and evaluation of the models. For MSI, we train on the colorectal cancer (CRC) cohort from The Cancer Genome Atlas (TCGA), TCGA-CRC, and evaluate on the CRC cohort from the Clinical Proteomic Tumor Analysis Consortium (CPTAC), CPTAC-CRC. For HRD, we train on the lung adenocarcinoma (LUAD) cohort from TCGA, TCGA-LUAD, and evaluate on the LUAD cohort from CPTAC, CPTAC-LUAD. The public biomarker data for MSI is from the study by Wagner et al.~\cite{Wagner2023-zn}, for HRD is from the study by El Nahhas et al.~\cite{El_Nahhas2024-xu}, and for the TME is from the study by Thorsson et al.~\cite{Thorsson2018-hi}. The slides for TCGA are available at \url{https://portal.gdc.cancer.gov}. The slides for CPTAC are available at \url{https://proteomics.cancer.gov/data-portal}. The overlap of patients with MSI or HRD status and the availability of TME targets for TCGA and CPTAC is found in Suppl.~Table~1.

\subsection{Joint multi-task learning improves classification predictions}
We develop a Transformer architecture that performs weakly-supervised classification and regression in a joint multi-task learning setting with \gls{wsi} features as input. To the best of our knowledge, there is no prior work that predicts either MSI or HRD directly from \glspl{wsi} in a joint multi-task setting. Therefore, we compare our work to the state-of-the-art MSI~\cite{Wagner2023-zn} and HRD~\cite{El_Nahhas2024-xu} weakly-supervised classification models, which have been trained and validated in large cohort studies. Since these models do not employ joint multi-task learning, we include an additional baseline where we employ our framework without the auxiliary regression task. Our baseline model outperforms the state-of-the-art MSI classification model with an AUROC of 86.1\% versus the reported AUROC of 83.0\%~\cite{Wagner2023-zn} in TCGA-CRC. For the prediction of HRD, our baseline model outperforms the state of the art with an AUROC of 71.6\% versus the reported AUROC of 70.0\%~\cite{El_Nahhas2024-xu} in TCGA-LUAD. When introducing auxiliary regression tasks to our model which learn to quantify the tumor microenvironment in a joint multi-task setting for the prediction of MSI and HRD, a substantial increase in performance is measured versus the state of the art and baseline performance. Specifically, adding auxiliary regression tasks to our model yields an AUROC of 94.0\% and AUPRC of 84.5\% for the prediction of MSI in TCGA-CRC, and an AUROC of 73.4\% and AUPRC of 59.8\% for the prediction of HRD in TCGA-LUAD (Table 1). This is an improvement over the state of the art by +11\% and +3.4\% in the respective cohorts as measured by the AUROC. These data show that weakly-supervised joint multi-task learning improves classification predictions over the baseline model and the state of the art.

\begin{table}[!ht]
\centering
\caption{Performance overview of weakly-supervised MSI and HRD biomarker prediction models.}
\label{table:1}
\begin{tabular}{l|cccc|cccc}
\hline
                  & \multicolumn{4}{c|}{MSI}                                                           & \multicolumn{4}{c}{HRD}                                                            \\ 
                  & \multicolumn{2}{c|}{TCGA-CRC}                         & \multicolumn{2}{c|}{CPTAC-CRC}    & \multicolumn{2}{c|}{TCGA-LUAD}                        & \multicolumn{2}{c}{CPTAC-LUAD}    \\ 
                  & AUC           & \multicolumn{1}{c|}{PRC}           & AUC           & PRC           & AUC           & \multicolumn{1}{c|}{PRC}           & AUC           & PRC           \\ \hline
SOTA              & 83.0          & \multicolumn{1}{c|}{-}             & 82.0          & -             & 70.0          & \multicolumn{1}{c|}{-}             & 82.0          & -             \\
baseline           & 86.1          & \multicolumn{1}{c|}{61.4}          & 86.4          & 70.5          & 71.6          & \multicolumn{1}{c|}{57.7}          & 81.0          & 30.3          \\ \hline
naive             & 86.4          & \multicolumn{1}{c|}{62.7}          & 88.2          & 72.4          & 69.6          & \multicolumn{1}{c|}{57.6}          & 81.2          & 33.7          \\
dwa               & 84.2          & \multicolumn{1}{c|}{58.7}          & 87.7          & 70.8          & 73.3          & \multicolumn{1}{c|}{60.3}          & 83.6          & 40.8          \\
uncert            & 86.0          & \multicolumn{1}{c|}{63.4}          & 88.4          & 72.6          & 73.2          & \multicolumn{1}{c|}{60.1}          & 83.1          & 41.5          \\
autol             & \textbf{94.0} & \multicolumn{1}{c|}{\textbf{84.5}} & 86.9          & 73.1          & 72.2          & \multicolumn{1}{c|}{58.5}          & 85.2          & 43.0          \\
graddrop          & 85.8          & \multicolumn{1}{c|}{61.2}          & 87.3          & 71.7          & 71.1          & \multicolumn{1}{c|}{58.8}          & 84.4          & 42.5          \\
pcgrad            & 85.4          & \multicolumn{1}{c|}{62.4}          & 87.3          & 72.0          & 72.1          & \multicolumn{1}{c|}{\textbf{60.4}} & 84.1          & 41.2          \\
cagrad            & 86.5          & \multicolumn{1}{c|}{62.7}          & 89.7          & \textbf{76.6} & 72.6          & \multicolumn{1}{c|}{58.7}          & 85.4          & 42.1          \\
dwa + graddrop    & 85.5          & \multicolumn{1}{c|}{59.8}          & 87.6          & 71.6          & 72.6          & \multicolumn{1}{c|}{58.4}          & 83.3          & 40.2          \\
dwa + pcgrad      & 85.6          & \multicolumn{1}{c|}{62.3}          & 88.8          & 73.3          & \textbf{73.4} & \multicolumn{1}{c|}{59.8}          & 83.0          & 39.1          \\
dwa + cagrad      & 85.8          & \multicolumn{1}{c|}{59.9}          & 88.4          & 73.8          & 71.8          & \multicolumn{1}{c|}{57.9}          & 85.5          & 44.3          \\
uncert + graddrop & 85.5          & \multicolumn{1}{c|}{60.7}          & 87.6          & 71.7          & 72.6          & \multicolumn{1}{c|}{59.4}          & 83.9          & 42.3          \\
uncert + pcgrad   & 86.7          & \multicolumn{1}{c|}{62.5}          & 89.0          & 74.2          & 71.8          & \multicolumn{1}{c|}{55.4}          & 83.6          & 39.9          \\
uncert + cagrad   & 86.3          & \multicolumn{1}{c|}{60.7}          & 88.9          & 74.6          & 72.4          & \multicolumn{1}{c|}{58.8}          & 84.4          & 41.9          \\
autol + graddrop  & 85.3          & \multicolumn{1}{c|}{61.6}          & 86.7          & 69.4          & 70.8          & \multicolumn{1}{c|}{56.0}          & 84.8          & 43.4          \\
autol + pcgrad    & 86.1          & \multicolumn{1}{c|}{63.2}          & 87.9          & 73.6          & 72.0          & \multicolumn{1}{c|}{58.2}          & 85.6          & 42.8          \\
autol + cagrad    & 86.5          & \multicolumn{1}{c|}{62.0}          & \textbf{89.9} & 76.3          & 71.9          & \multicolumn{1}{c|}{57.0}          & \textbf{86.1} & \textbf{43.8} \\ \hline
\end{tabular}
\end{table}

\subsection{Joint multi-task learning improves generalizability}
Next, we evaluate the generalizability of the joint multi-task learned models' performance to external cohorts and compare them to the baseline and the state of the art in MSI~\cite{Wagner2023-zn} and HRD~\cite{El_Nahhas2024-xu} classification (Table 1). The baseline model outperforms the state-of-the-art MSI model performance by +4.4\% in the AUROC metric, whereas the baseline HRD model yields slightly inferior AUROCs by -1\%. Again, introducing auxiliary regression tasks to the model with weighting- and gradient-balancing schemes substantially improves the performance on external cohorts for prediction of MSI and HRD. The model with \textit{autol + cagrad} balancing yields an AUROC of 89.9\% and AUPRC of 76.3\% in predicting MSI in CPTAC-CRC, and an AUROC of 86.1\% and AUPRC of 43.8\% in predicting HRD in CPTAC-LUAD. This is a +7.7\% and +4.1\% improvement over the state-of-the-art models for MSI and HRD prediction in external cohorts, respectively. Notably, all combinations of joint multi-task learning using weighting- and gradient-based balancing, except for naive balancing, yield better AUROCs across the tested cohorts and targets. This underlines the need for the application of sophisticated multi-task balancing methods, which are neglected in prior work. Together, these data show that weakly-supervised joint multi-task learning with biologically relevant auxiliary regression tasks of the TME improves the prediction performance of key predictive biomarkers like MSI and HRD on external cohorts.

\subsection{Joint multi-task learning improves latent-embedding clustering}
Finally, we analyze the latent space of the classification head input~(\cref{fig:overview}) in both the classification and joint-learning setting. Measuring the clustering capabilities of the \textit{384}-dimensional embeddings through the SS, the best clustering performance is observed in the joint-learned embeddings (Table 2). Specifically, the best weighting- and gradient-based balancing combination for MSI is \mbox{(\textit{autol~+~cagrad})} with an SS of 0.44, and for HRD is \mbox{(\textit{dwa~+~cagrad})} and \mbox{(\textit{uncert~+~cagrad})} with an SS of 0.12. This is a +8\% increase over the mean clustering performance of the baseline for MSI, and +5\% for HRD. Interestingly, all models using weighting-based balancing together with \textit{cagrad} yield the best embedding clusters. Moreover, we visualize the \textit{384}-dimensional embeddings in 2D using t-SNE (Suppl.~Fig.~2), showing an equal AUC of 87\% between the baseline MSI-embeddings and joint-learned (MSI~+~Prolif)-embeddings, but yielding a substantially improved SS for the joint-learned embeddings (0.52 versus 0.33) in CPTAC-CRC. These findings collectively demonstrate that our proposed model improves the latent-embedding clustering performance in an external cohort, again highlighting improved generalizability over the baseline.

\begin{table}[!ht]

\centering
\caption{Clustering performance of the latent embeddings on external cohorts as measured by the silhouette score.}
\label{table2}
\begin{tabular}{c|cccccc|ccc}
\hline
\multicolumn{1}{l|}{} & \multicolumn{6}{c|}{\begin{tabular}[c]{@{}c@{}}MSI \\ CPTAC-CRC\end{tabular}} & \multicolumn{3}{c}{\begin{tabular}[c]{@{}c@{}}HRD \\ CPTAC-LUAD\end{tabular}} \\ \hline
Methods               & ITH            & LF             & LISS & Prolif        & SF            & mean & ITH                      & Prolif                   & mean                    \\ \hline
baseline              & 0.35           & 0.34           & 0.38 & 0.35          & 0.38          & 0.36 & 0.05                     & \textbf{0.10}            & 0.07                    \\ \hline
naive                 & 0.43           & 0.39           & 0.42 & 0.41          & \textbf{0.45} & 0.42 & 0.05                     & 0.08                     & 0.08                    \\
dwa                   & 0.26           & 0.33           & 0.33 & 0.33          & 0.27          & 0.30 & 0.07                     & 0.01                     & 0.04                    \\
uncert                & 0.30           & 0.34           & 0.39 & 0.33          & 0.31          & 0.33 & 0.07                     & 0.01                     & 0.04                    \\
autol                 & 0.37           & \textbf{0.48}           & 0.41 & 0.37          & 0.39          & 0.40 & 0.10                     & 0.07                     & 0.09                    \\
graddrop              & 0.31           & 0.35           & 0.29 & 0.32          & 0.27          & 0.31 & 0.08                     & 0.04                     & 0.06                    \\
pcgrad                & 0.28           & 0.44           & 0.38 & 0.32          & 0.35          & 0.35 & 0.08                     & 0.01                     & 0.05                    \\
cagrad                & 0.36           & \textbf{0.48}  & 0.39 & \textbf{0.45}          & 0.43          & 0.42 & 0.14                     & 0.05                     & 0.10                    \\
dwa + graddrop        & 0.31           & 0.31           & 0.33 & 0.31          & 0.27          & 0.31 & 0.04                     & 0.03                     & 0.04                    \\
dwa + pcgrad          & 0.31           & 0.39           & 0.38 & 0.38          & 0.30          & 0.35 & 0.05                     & 0.03                     & 0.04                    \\
dwa + cagrad          & 0.33           & 0.40           & \textbf{0.45} & 0.43          & 0.43          & 0.41 & 0.15                     & 0.08                     & \textbf{0.12}           \\
uncert + graddrop     & 0.31           & 0.31           & 0.32 & 0.31          & 0.23          & 0.30 & 0.08                     & 0.03                     & 0.06                    \\
uncert + pcgrad       & 0.33           & 0.44           & 0.34 & 0.38          & 0.33          & 0.36 & 0.05                     & 0.02                     & 0.04                    \\
uncert + cagrad       & 0.37           & \textbf{0.48}           & 0.44 & 0.38          & 0.43          & 0.42 & \textbf{0.16}            & 0.07                     & \textbf{0.12}           \\
autol + graddrop      & 0.32           & 0.30           & 0.29 & 0.32          & 0.29          & 0.30 & 0.10                     & 0.08                     & 0.09                    \\
autol + pcgrad        & 0.34           & 0.37           & 0.31 & 0.35          & 0.32          & 0.34 & 0.10                     & 0.06                     & 0.08                    \\
autol + cagrad        & \textbf{0.44}  & 0.45           & 0.43 & \textbf{0.45} & 0.41          & \textbf{0.44} & 0.12            & 0.07                     & 0.10                    \\ \hline
\end{tabular}
\end{table}

\section{Conclusion}
We have developed a weakly-supervised joint multi-task Transformer architecture which learns additional biological information from the tumor microenvironment to improve the prediction of MSI and HRD. Whereas existing research in computational pathology used naive multi-task balancing approaches, this study emphasizes the application of more sophisticated, model-guided balancing approaches which adapt to the weakly-supervised multi-task problem at hand. We conducted an ablation study of 16 weighting- and gradient-based multi-task balancing approaches, showing task balancing substantially impacts joint multi-task performance in weakly-supervised computational pathology. Our proposed approach yields state of the art performance in the weakly-supervised task of classifying MSI and HRD directly from \glspl{wsi} in 2 patient cohorts, as well as improved generalizability to 2 external patient cohorts. Moreover, we demonstrate that weakly-supervised joint multi-task learning with an auxiliary regression tasks improves the clustering capability of the latent space embedding. This work underlines the potential of biology-informed deep learning using auxiliary regression tasks to improve the main classification objective for highly relevant predictive biomarkers. Summarizing, we provide an open-source, weakly-supervised multi-task learning framework in computational pathology which jointly learns classification and regression tasks to outperform state-of-the-art classification models, trained and evaluated on publicly available patient cohorts.

\clearpage


\bibliographystyle{splncs04}
\bibliography{paperpile}

\begin{thebibliography}{10}
\providecommand{\url}[1]{\texttt{#1}}
\providecommand{\urlprefix}{URL }
\providecommand{\doi}[1]{https://doi.org/#1}

\bibitem{Bai2021-wg}
Bai, J., Chen, H., Bai, X.: Relationship between microsatellite status and immune microenvironment of colorectal cancer and its application to diagnosis and treatment. J. Clin. Lab. Anal.  \textbf{35}(6),  e23810 (Jun 2021)

\bibitem{Campanella2019-ni}
Campanella, G., et~al.: Clinical-grade computational pathology using weakly supervised deep learning on whole slide images. Nat. Med.  \textbf{25}(8),  1301--1309 (Aug 2019)

\bibitem{Chen2020-fd}
Chen, Z., Ngiam, J., Huang, Y., Luong, T., Kretzschmar, H., Chai, Y., Anguelov, D.: Just pick a sign: Optimizing deep multitask models with gradient sign dropout  (Oct 2020)

\bibitem{Dosovitskiy2020-gr}
Dosovitskiy, A., et~al.: An image is worth 16x16 words: Transformers for image recognition at scale (Oct 2020)

\bibitem{El_Nahhas2023-du}
El~Nahhas, O.S.M., et~al.: From whole-slide image to biomarker prediction: A protocol for {End-to-End} deep learning in computational pathology  (Dec 2023)

\bibitem{El_Nahhas2024-xu}
El~Nahhas, O.S.M., et~al.: Regression-based {Deep-Learning} predicts molecular biomarkers from pathology slides. Nat. Commun.  \textbf{15}(1),  1--13 (Feb 2024)

\bibitem{Gao2023-no}
Gao, Z., et~al.: A semi-supervised multi-task learning framework for cancer classification with weak annotation in whole-slide images. Med. Image Anal.  \textbf{83},  102652 (Jan 2023)

\bibitem{Graham2023-iv}
Graham, S., et~al.: One model is all you need: Multi-task learning enables simultaneous histology image segmentation and classification. Med. Image Anal.  \textbf{83},  102685 (Jan 2023)

\bibitem{Kather2019-jh}
Kather, J.N., et~al.: Deep learning can predict microsatellite instability directly from histology in gastrointestinal cancer. Nat. Med.  \textbf{25}(7),  1054--1056 (Jun 2019)

\bibitem{Kendall2017-vx}
Kendall, A., Gal, Y., Cipolla, R.: {Multi-Task} learning using uncertainty to weigh losses for scene geometry and semantics  (May 2017)

\bibitem{Liu2021-cs}
Liu, B., Liu, X., Jin, X., Stone, P., Liu, Q.: {Conflict-Averse} gradient descent for multi-task learning  (Oct 2021)

\bibitem{Liu2022-ao}
Liu, S., James, S., Davison, A.J., Johns, E.: {Auto-Lambda}: Disentangling dynamic task relationships  (Feb 2022)

\bibitem{Liu2018-rh}
Liu, S., Johns, E., Davison, A.J.: {End-to-End} {Multi-Task} learning with attention  (Mar 2018)

\bibitem{Loeffler2023-vt}
Loeffler, C.M.L., et~al.: Direct prediction of homologous recombination deficiency from routine histology in ten different tumor types with attention-based multiple instance learning: a development and validation study. medRxiv  (Mar 2023)

\bibitem{Loshchilov2018-uv}
Loshchilov, I., Hutter, F.: Decoupled weight decay regularization (Sep 2018)

\bibitem{Lu2021-pp}
Lu, M.Y., et~al.: {AI-based} pathology predicts origins for cancers of unknown primary. Nature  \textbf{594}(7861),  106--110 (Jun 2021)

\bibitem{Marini2021-em}
Marini, N., et~al.: {Multi-Scale} task multiple instance learning for the classification of digital pathology images with global annotations. In: Proceedings of the {MICCAI} Workshop on Computational Pathology. Proceedings of Machine Learning Research, vol.~156, pp. 170--181. PMLR (Sep 2021)

\bibitem{Mormont2020-zs}
Mormont, R., Geurts, P., Mar{\'e}e, R.: {Multi-Task} {Pre-Training} of deep neural networks for digital pathology. IEEE journal of biomedical and health informatics  (2020)

\bibitem{Niehues2023-ew}
Niehues, J.M., et~al.: Generalizable biomarker prediction from cancer pathology slides with self-supervised deep learning: A retrospective multi-centric study. Cell Rep Med p. 100980 (Mar 2023)

\bibitem{Shi2023-pf}
Shi, Z., Chen, B., Han, X., Gu, W., Liang, S., Wu, L.: Genomic and molecular landscape of homologous recombination deficiency across multiple cancer types. Sci. Rep.  \textbf{13}(1), ~8899 (Jun 2023)

\bibitem{Tellez2020-gz}
Tellez, D., et~al.: Extending unsupervised neural image compression with supervised multitask learning. In: Proceedings of the Third Conference on Medical Imaging with Deep Learning. Proceedings of Machine Learning Research, vol.~121, pp. 770--783. PMLR (2020)

\bibitem{Thorsson2018-hi}
Thorsson, V., et~al.: The immune landscape of cancer. Immunity  \textbf{48}(4),  812--830.e14 (Apr 2018)

\bibitem{Vaswani2017-ht}
Vaswani, A., et~al.: Attention is all you need. In: Advances in Neural Information Processing Systems. vol.~30. Curran Associates, Inc. (2017)

\bibitem{Wagner2023-zn}
Wagner, S.J., et~al.: Transformer-based biomarker prediction from colorectal cancer histology: A large-scale multicentric study. Cancer Cell  \textbf{41}(9),  1650--1661.e4 (Sep 2023)

\bibitem{Wang2022-wg}
Wang, X., et~al.: Transformer-based unsupervised contrastive learning for histopathological image classification. Med. Image Anal.  \textbf{81},  102559 (Oct 2022)

\bibitem{Wolflein2023-vp}
W{\"o}lflein, G., et~al.: A good feature extractor is all you need for weakly supervised learning in histopathology  (Nov 2023)

\bibitem{Yan2020-dg}
Yan, C., Xu, J., Xie, J., Cai, C., Lu, H.: {Prior-Aware} {CNN} with {Multi-Task} learning for colon images analysis. In: 2020 {IEEE} 17th International Symposium on Biomedical Imaging ({ISBI}). pp. 254--257. IEEE (Apr 2020)

\bibitem{Yu2020-wy}
Yu, T., Kumar, S., Gupta, A., Levine, S., Hausman, K., Finn, C.: Gradient surgery for {Multi-Task} learning  (Jan 2020)

\end{thebibliography}
\clearpage

\subsubsection{Acknowledgements.}
This research has been partially funded by the German Federal Ministry of Education and Research (BMBF) through grant 1IS23070, Software Campus 3.0 (TU Dresden), as part of the Software Campus project 'MIRACLE-AI'. JNK is supported by the German Federal Ministry of Health (DEEP LIVER, ZMVI1-2520DAT111), the German Cancer Aid (DECADE, 70115166), the German Federal Ministry of Education and Research (PEARL, 01KD2104C; CAMINO, 01EO2101; SWAG, 01KD2215A; TRANSFORM LIVER, 031L0312A; TANGERINE, 01KT2302 through ERA-NET Transcan), the German Academic Exchange Service (SECAI, 57616814), the German Federal Joint Committee (TransplantKI, 01VSF21048) the European Union’s Horizon Europe and innovation programme (ODELIA, 101057091; GENIAL, 101096312), the European Research Council (ERC; NADIR, 101114631) and the National Institute for Health and Care Research (NIHR, NIHR213331) Leeds Biomedical Research Centre. The views expressed are those of the author(s) and not necessarily those of the NHS, the NIHR or the Department of Health and Social Care. This work was funded by the European Union. Views and opinions expressed are however those of the author(s) only and do not necessarily reflect those of the European Union. Neither the European Union nor the granting authority can be held responsible for them.

\subsubsection{Competing interests.}
OSMEN holds shares in StratifAI GmbH. FK holds shares in StratifAI GmbH. DT holds shares in StratifAI GmbH. JNK declares consulting services for Owkin, France, DoMore Diagnostics, Norway, Panakeia, UK, Scailyte, Switzerland, Cancilico, Germany, Mindpeak, Germany, MultiplexDx, Slovakia, and Histofy, UK; furthermore he holds shares in StratifAI GmbH, Germany, has received a research grant by GSK, and has received honoraria by AstraZeneca, Bayer, Eisai, Janssen, MSD, BMS, Roche, Pfizer and Fresenius. The mentioned competing interests are related to cancer and the computational analysis of histopathology slides, which is the main topic of this research.

\clearpage

\renewcommand{\figurename}{Suppl. Fig.}
\renewcommand{\tablename}{Suppl. Table}
\renewcommand{\thesection}{\Alph{section}}
\setcounter{figure}{0}
\setcounter{table}{0}
\setcounter{section}{0}
\section{Supplementary material}
\begin{figure}
    \centering
    \includegraphics[width=0.85\textwidth,height=\textheight,keepaspectratio]{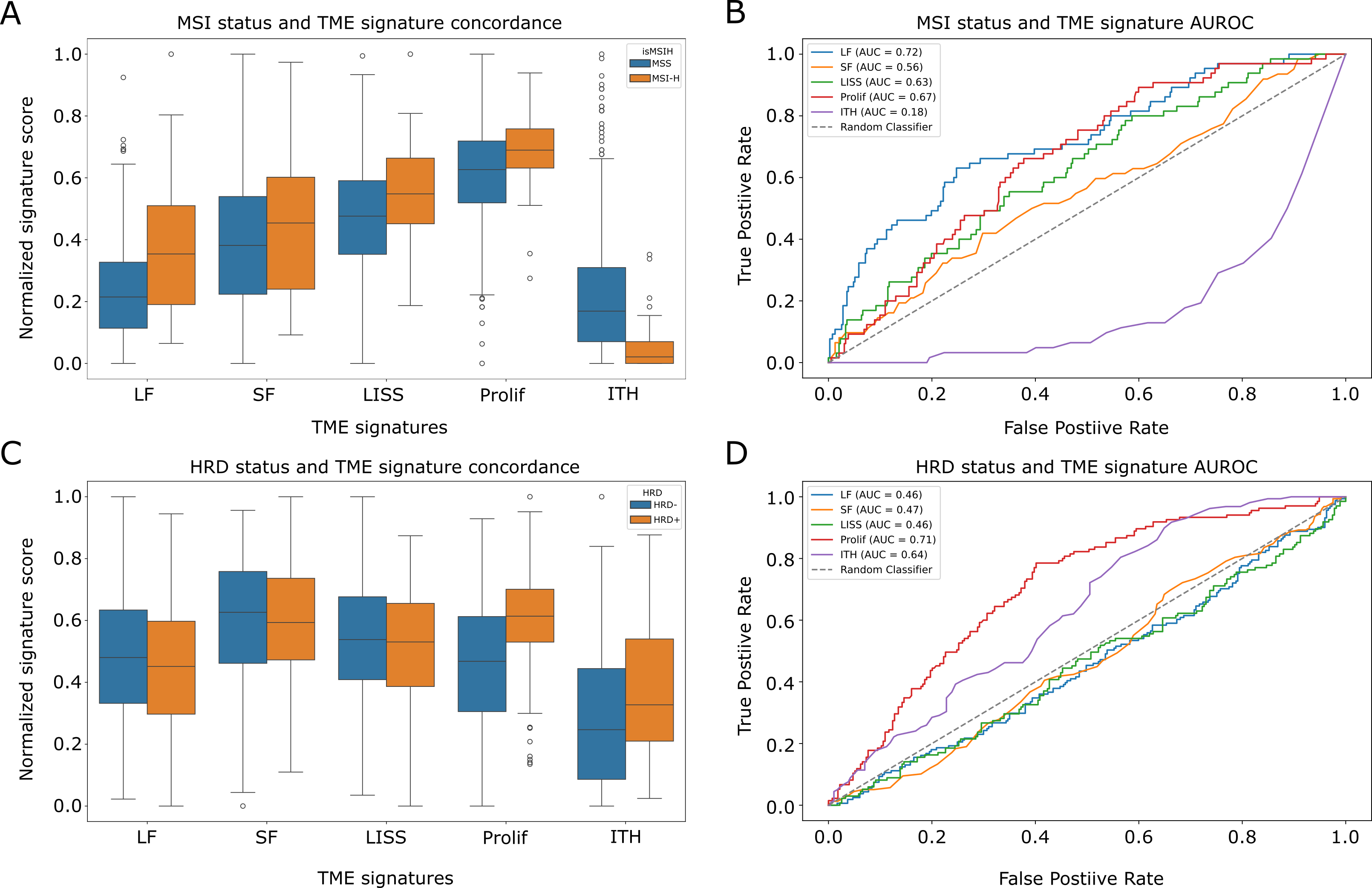}
    \caption{The relationship between TME signatures, and \textbf{a,b} MSI and \textbf{c,d} HRD. TME signatures with non-random relationships with MSI and HRD were used for subsequent experiments.}
    \label{fig:enter-label}
\end{figure}

\begin{table}[]
\centering
\label{}
\caption{Data overview for training and validation, number of patients. The TME information is not available for CPTAC and is thus blindly deployed to measure prediction performance of MSI and HRD.}
\begin{tabular}{l|cc}
\hline
             & TCGA & CPTAC \\ \hline
MSI U LISS   & 427  & 105   \\
MSI U ITH    & 421  & 105   \\
MSI U Prolif & 427  & 105   \\
MSI U SF     & 421  & 105   \\
MSI U LF     & 427  & 105   \\ \hline
HRD U ITH    & 433  & 106   \\
HRD U Prolif & 400  & 106   \\ \hline
\end{tabular}
\end{table}

\begin{figure}[]
    \centering
    \includegraphics[width=0.85\textwidth,height=\textheight,keepaspectratio]{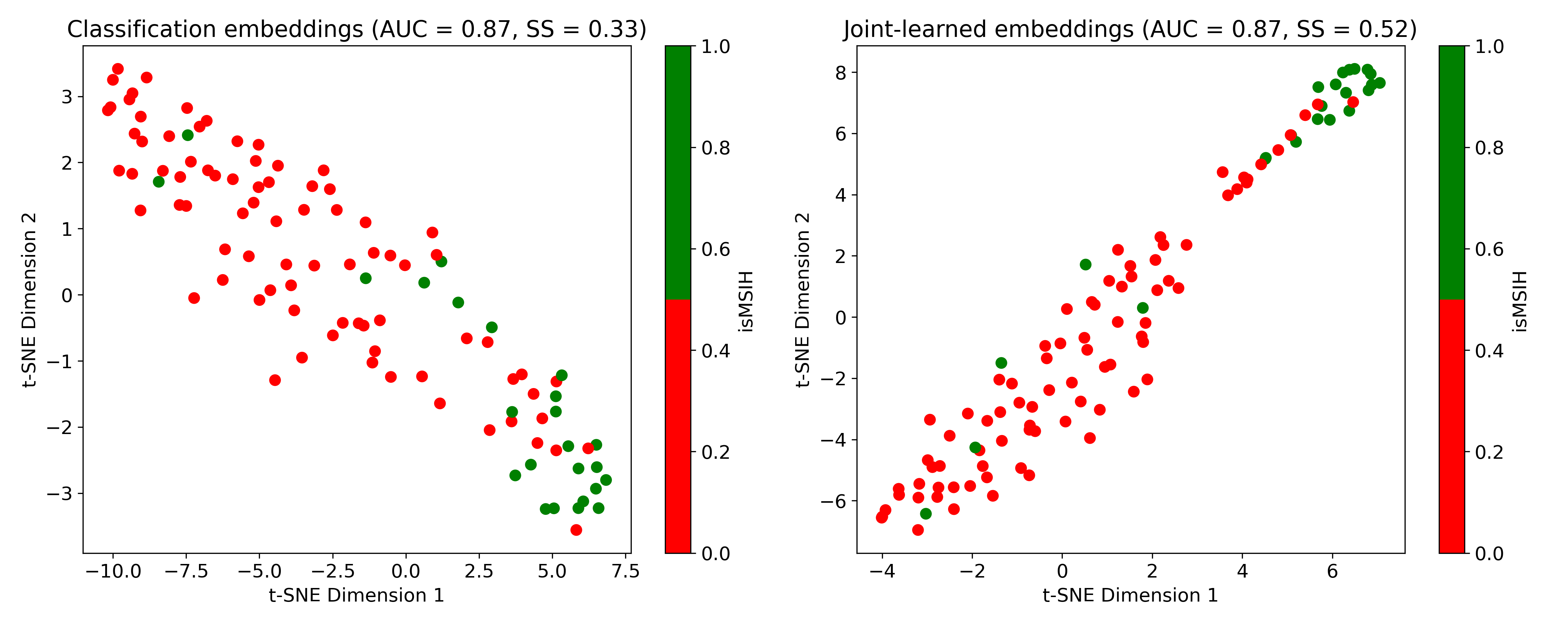}
    \caption{Visualization of the classification and joint-learned embeddings for MSI of the external cohort (n=105) using t-SNE}
    \label{fig:fig2}
\end{figure}

\end{document}